\documentclass[prl,aps,showpacs,twocolumn]{revtex4}
\usepackage{epsfig}
\usepackage{bm}

\begin{document}

\title{Towards a dynamical theory of multifractals in turbulence}
\author{Victor Yakhot$^{1,2}$ and K. R. Sreenivasan$^{2}$\\
$^1$Department of Aerospace and Mechanical Engineering,\\ Boston
University, Boston 02215\\ $^2$International Center for
Theoretical Physics, Trieste, Italy}

\begin{abstract} \noindent  Making use of the exact
equations for structure functions, supplemented by the equations for dissipative
anomaly as well as an estimate for the Lagrangian acceleration of fluid
particles, we obtain a main result of the multifractal theory of
turbulence. The central element of the theory is a dissipation
cut-off that depends on the order of the structure function. An
expression obtained for the exponents $s_{n}$ in the scaling
relations $\overline{(\frac{\partial u}{\partial
x})^{n}}/\overline{(\frac{\partial u}{\partial
x})^{2}}^{\frac{n}{2}} \propto Re^{s_{n}}$, between the velocity
gradients $\frac{\partial u}{\partial x}$ and the Reynolds number
$Re$, agrees well with experimental data.
\end{abstract}


\maketitle

Questions of small-scale universality in fluid turbulence hover
around the universality of the scaling exponents $\xi_{n,0}$ of
velocity structure functions defined through relations such as
\begin{equation}
S_{n,0}=\overline{[u(x+r)-u(x)]^n}\equiv
\overline{(\delta_{r}u)^{n}}\propto (\overline \epsilon
L)^{\frac{n}{3}}(\frac{r}{L})^{\xi_{n,0}},
\end{equation}
where $u(x)$ is the velocity component along the separation
distance $r$, measured at the position $x$ and $\overline
\epsilon$ is the mean rate of energy dissipation. Here $r$ lies in
the inertial range given by $\eta << r << L$, where $L$ is the
large-scale at which the energy is being injected and $\eta \equiv
(\nu^3/\overline \epsilon)^{1/4}$ is the dissipation scale, $\nu$
being the fluid viscosity. The zero index in $\xi_{n,0}$ shows
that no powers of the transverse velocity increments are involved
in this particular definition (1). Kolmogorov [1] assumed that the
velocity fluctuations in the inertial range are independent of
both $L$ and $\eta$, and that $\overline \epsilon$, regarded as
equal to the energy flux across scales, is the only relevant
dynamical parameter. As is well known, Kolmogorov's proposal
yields the linear relation $\xi_{n,0}=n/3$. Since, in the limit of
vanishing viscosity (or, as $\eta \rightarrow 0$), Kolmogorov's
scaling theory combines the exact expression [2] $S_{3,0}=
-\frac{4}{5}\overline {\epsilon}r$, it is reasonable to regard the
theory loosely as dynamic. However, experimental and numerical
data in three-dimensional turbulence have shown (see Ref.\ [3] for a recent account)
that the scaling exponents $\xi_{n,0}$
depart from $n/3$, and that there exists a more complicated
nonlinear spectrum of scaling exponents $\xi_{n,0}$. Its
theoretical explanation for the velocity field has proved to be
elusive, though considerable progress has been made for passive
scalars [4].

In recent past, the problem of scaling exponents in turbulence has
been analyzed within a general framework of the theory of
multifractal (MF) processes reviewed in
Ref.\ [5]. This approach has led to interesting interpretations
and novel work (see [5,6] for incomplete list), but its
shortcoming is the lack of connection with the dynamical
equations. In this paper, a main relation of the MF theory is
derived from dynamical equations, supplemented both by an
order-of-magnitude estimate for the Lagrangian acceleration of a
fluid particle, and the earlier work on dissipative anomaly
[7,8].

For background, we review here the main ideas of the
inertial-range MF theory, whose basis are the assumptions that (a)
the velocity increments $\delta_{r}u$ have the form
\begin{equation}
\frac{\delta_{r} u({\bf x})}{u^{\prime}} \equiv \frac{u({\bf
x+r})-u({\bf x})}{u^{\prime}} \propto (\frac{r}{L})^{h},
\end{equation}
where $u^{\prime} \sim \delta_{L} u$ may be regarded as the
root-mean-square value of $u$, and (b) there exists a spectrum of
exponents $h$ related to the fractal dimension of their support
$D(h)$. Thus, $(r/L)^{3-D(h)}$ is proportional to the probability
of the velocity increment falling within a sphere of radius $r$ on
a set of dimension $D(h)$. It is clear from (2) that
\begin{equation}
S_{n,0}=\overline{(\delta_{r} u)^{n}} \propto
(\frac{r}{L})^{\xi_{n,0}}= \int d\mu(h)(\frac{r}{L})^{nh+3-D(h)},
\end{equation}
where $d\mu(h)$ is the weight of a local value of exponent $h$.
Thus the scaling exponents $\xi_{n,0}$ are directly related to the
spectrum $D(h)$. The goal of the theory is to find the
functions $D(h)$ and $\mu(h)$. If one evaluates the integral in
(3) in the steepest descent approximation, as in the standard
procedure, the precise form of $\mu(h)$ is irrelevant and only the
spectrum $D(h)$ needs to be determined. However, this cannot be
done within the MF theory itself.

Multifractality in the inertial range will have consequences for
dissipation scales as well. The authors of Ref.\ [9] used the
local scaling (2) to construct eddy-turnover times that depended
on $h$, equated them to diffusion times scales $\eta^2/\nu$, and
showed that a spectrum of $h$-dependent dissipation scales can be
written as
\begin{equation}
\eta(h) \propto LRe^{-\frac{1}{1+h}},
\end{equation}
where the large-scale Reynolds number $Re = u^{\prime}L/\nu$. The
exponents in (4) have to be related somehow to the spectrum of
scaling exponents of structure functions. This can be done by
assuming, for any $h$, that (2) is valid only for scales $r >
\eta(h)$, with smoothness for smaller scales [10,5]. One can then
evaluate (3) for scales larger than $\eta(h)$,
using the steepest descent approximation up to the cut-off scale
$r=\eta(h)$. It is easy to show [5] that
\begin{equation}
\overline{(\partial_{x}u)^{n}} \propto
Re^{\zeta_{n,0}+\frac{n}{2}} \equiv Re^{s_n},
\end{equation}
where $\zeta_{n,0} = p(n,0)-\frac{3n}{2}$ and $p(n,0)$ is the
solution of $p(n,0) = 2n - \xi_{p,0}$. Our specific goal is to
obtain $s_n$ theoretically.

A brief remark on our strategy may be helpful here. If the
structure functions $S_{n,m} \equiv \overline{(\delta_{r}
u)^{n}(\delta_{r} v)^{m}}$, where $v$ now is the velocity
component normal to the displacement vector ${\bf r}$, of the form
$A_{n,m}r^{\xi_{n,m}}$, are non-analytic in the inertial range,
and the viscous dissipation is the only mechanism for smoothing
the singular nature of the structure functions in the inertial
range, the balance between them occurs at the length scale $r
\rightarrow \eta_{n,0}$, where $\eta_{n,0} \equiv \eta_{n}$ is an
order-dependent length scale nominally separating the analytic and
singular intervals. The analyticity of structure functions in the
viscous range yields $S_{n,0}\propto \overline{(\partial_{x}
u(0))^{n}}r^{n}$, so we have [11]
\begin{equation}
\eta_{n}\approx (\overline \epsilon
L)^{\frac{n}{(3(n-3))}}\overline{(\partial_{x}
u)^{n}}^{\frac{1}{(\xi_{n,0})-n}}\approx \overline{(\partial_{x}
u)^{n}}^{\frac{1}{(\xi_{n,0})-n}}.
\end{equation}
This equation defines the field $\eta({\bf x},t)$ through moments
of velocity gradients, and picks out the strongest singularity of
a chosen order dominating the inertial range asymptotics. Our
strategy is based on the idea that if an $n$-th order structure
function evaluated at the appropriate cut-off is
$S_{n,0}(\eta_{n,0})=A_{n}\eta^{\xi_{n,0}}$, with the
Reynolds-number-independent proportionality coefficient $A_{n}$,
then $S_{n,0}(r)\propto r^{\xi_{n,0}}$ for $r$ in the inertial
range. (Henceforth, to simplify notation, we will often set
$\overline \epsilon=L=1$ and omit the second index in the $\xi$'s
and $\eta$'s.)

As the first step in the theory, we write the exact equation for
structure function of order $2n$ [12,11] (see also Refs.\
[13,3,14]) as
\begin{eqnarray}
\frac{\partial S_{2n,0}}{\partial
r}+\frac{d-1}{r}S_{2n,0}=\frac{(d-1)(2n-1)}{r}S_{2n-2,2} \nonumber\\
+(2n-1)\overline{\delta_{r}a_x U^{2n-2}},
\end{eqnarray}
where the increment of the $r$-component of Lagrangian
acceleration of a fluid particle is given by
\begin{equation}
\delta_{r}a_x=-[\partial_{x'}p(x')-\partial_{x}p(x)]+\nu
[\nabla_{{\bf x'}}^{2}u({\bf x^{\prime}})-\nabla_{\bf x}^{2}u({\bf
x})]
\end{equation}
and ${\bf x'}={\bf x+r}$.

The second step requires the closure of Eq.\ (7), for which we
need an expression for $\delta_{r}a_x$ in terms of velocity
increments. The Laplacian in (8) can be represented in terms of
finite differences on the dissipation cut off $\eta$ and, by
virtue of Eq.\ (6), the acceleration increment can be made a
function of two fluctuating variables (operators) $\delta_{\eta}
u$ and $\eta$. To make further progress, however, it is necessary
to express $\eta$ in terms of the velocity field itself.

We consider two scenarios. In the first, we have the option of
expressing the acceleration terms through either a model for the
conditional mean of the acceleration increment for a fixed value
of the velocity increment, or through a direct relation between
$\delta a$ and $\delta u$---somewhat in the spirit of Kolmogorov's
refined similarity hypothesis [15]. Choosing the former option, we
model in the limit $r\rightarrow \eta$ where $\eta$ is a generic
local dissipation scale, the $x$-component of acceleration term as
[16]
\begin{equation}
\overline{\delta_{\eta}a_x|\delta_{\eta}u} \approx
\overline{\frac{\delta_{\eta} u}{\tau}|\delta_\eta u} \approx
\overline{\frac{(\delta_{\eta}
u)^{2}}{\eta}|\delta_{\eta}u}\approx
\frac{(\delta_{\eta}u)^{3}}{\nu},
\end{equation}
where $\tau\approx \eta/\delta_{\eta} u$ is the life-time of a
fluctuation on the scale $\eta$. In the last step in Eq.\ (9), we
have used
\begin{equation}
\nu\approx \eta\delta_{\eta}u,
\end{equation}
where $\eta$ is to be regarded as a random field. This step
reduces the number of random fields from 2 to 1. Expression (10),
which is central for our theory, is proposed here on dimensional
grounds but will be obtained below from a second scenario
considering dissipative anomaly.

This second scenario follows Polyakov's work [7] on statistically
steady turbulence due to the one-dimensional Burgers equation
stirred by a large-scale random force. In that work, on the basis
of the energy balance equation for $\nu=0$, namely,
\begin{equation}
\frac{1}{2}\frac{\partial u^{2}}{\partial
t}+\frac{1}{3}\frac{\partial u^{3}}{\partial x}=fu,
\end{equation}
Polyakov derived the dissipation anomaly, which is related to the
local form of the Kolmogorov law [2] as
\begin{eqnarray}
-\frac{d}{dt}(u(x+\frac{y}{2})u(x-\frac{y}{2}))\approx
\frac{2}{3}\frac{\partial}{\partial x}u^{3}+ \nonumber \\
\lim_{y\rightarrow \eta}\frac{1}{6}\frac{\partial}{\partial
y}(u(x+\frac{y}{2})-u(x-\frac{y}{2}))^{3}=D.
\end{eqnarray}
Here, $y\rightarrow \eta\rightarrow 0$, and $D\approx -F =
f(x+\frac{y}{2})u(x-\frac{y}{2})+f(x-\frac{y}{2})u(x+\frac{y}{2})$
when $\nu = 0$, while $D= \nu
(u(x+y)\partial_{x-y}^{2}u(x-y)+u(x-y)\partial^{2}_{x+y}u(x+y))$
when the forcing $f$ is zero. Equation (12) balances the singular
contributions in the limit $y\rightarrow\eta\rightarrow 0$, while
the regular contributions disappear by virtue of (11). The
coordinate shift $y$ in Eq.\ (12) is identical  to Kolmogorov's
displacement $r=y\rightarrow\eta$. If, as $\eta\rightarrow 0$, the
velocity field is non-differentiable (i.e., singular), the left
side of Eq.\ (12) does not approach zero even in the limit
$\nu\rightarrow 0$. In a statistically steady state, Eq.\ (12)
immediately gives $\overline{(u(x+y)-u(x))^{3}}=-12 Fy$ for the
inertial range $y=r<<L$. We can see that the celebrated $-4/5$-ths
law of Kolmogorov [2] is not locally valid because of the
$O(\partial_{x}u^{3})$ term in (12); this term can, however, be
eliminated by averaging (12) over the directions of
velocity vector ${\bf u}/u$.

Now using the finite difference definition of all derivatives on a
dissipation scale $\eta\rightarrow 0$, we can write, after some
algebra, that
\begin{eqnarray}
\frac{1}{3}\frac{u^{3}(x^{+})-u^{3}(x^{-})}{\eta}+\frac{(u(x+2\eta)-u(x))^{3}}{6\eta} \nonumber \\
\approx [2\nu/\eta^2] \times [u^{2}(x^{+})+u^{2}(x^{-})+ u(x^{+})u(x-3\eta) \nonumber \\
+ u(x^{-})u(x+3\eta)-4u(x^{+})u(x^{-})],
\end{eqnarray}
where $x^{\pm}=x\pm\eta$. This equation is correct up to
$O(\eta^{2})$. As mentioned earlier, the single-point contribution
to this relation disappears when averaged over the ``directions''
of $\eta$. While the left side of (13) involves two-point
differences, the right side includes contributions from four
shifted points. To proceed further, we assume that $\eta$ plays
the same role as the width of typical shock structures, and
conclude that $u(x+3\eta)-u(x)\approx u(x^{+})-u(x^{-})\approx
u(x+2\eta)-u(x)$, as a result of which the right side of (13) is
$O[\nu (\delta_{\eta} u)^2/\eta^{2}]$. This leads to (10).

In three dimensions, however, additional terms appear due to the
pressure gradient-velocity product. The relevant extensions have
been made in Refs.\ [8]. The finite-difference representation of
the equations from Refs.\ [8] on the dissipation scale
$\eta\rightarrow 0$ yield the estimate
\begin{equation}
\frac{(\delta_{\eta} u)^{3}}{\eta}\approx
-\delta_{\eta}(\frac{\partial p}{\partial x}u)-
\nu\frac{(\delta_{\eta} u)^{2}}{\eta^{2}}.
\end{equation}
Since on the dissipation scale the pressure and dissipation
contributions are of the same order [16], expression (14) gives
the same balance relation, $\nu\approx \eta\delta_{\eta}u$,
obtained above. Thus, the relation (10) applies also to
three-dimensional turbulence.

Two comments are in order. First, the model for acceleration
should include the $O((\delta_{\eta} u)^{2}/\eta)$ quadratic
contribution coming from the pressure terms [16]. However, the
pressure term simply renormalizes the coefficients in front of the
remaining contributions to (7) and, as a result, does not alter
the steps presented above (see also Ref.\ [11]). Second, to our knowledge, there
are no experimental or numerical data that directly address the
conditional acceleration term of Eq.\ (9), though related
conditional data are accumulating rapidly [17,18].

Substituting Eq.\ (9) (after using Eq.\ (10)) into Eqs.\ (7) and
(8) we obtain an infinite set of equations coupling the structure
functions $S_{2n}(r)$ and $S_{2n+1}(r)$. These equations are valid
for all magnitudes of displacement $r\ll L$, including
$r\rightarrow \eta_{2n}$. It will become clear below that
$\eta_{2n}\geq \eta_{2n+1}$ and, as a result both
$S_{2n}(\eta_{2n})$ and $S_{2n+1}(\eta_{2n})$ are in their
respective algebraic ranges, i.e., $S_{2n}(\eta_{2n})\propto
\eta_{2n}^{\xi_{2n}}$ and $S_{2n+1}(\eta_{2n})\propto
\eta_{2n}^{\xi_{2n+1}}$. Thus, on the scale $\eta_{2n}$, Eqs.\ (9)
and (7) give
\begin{equation}
\eta_{2n}^{\xi_{2n}-1}\approx Re \eta_{2n}^{\xi_{2n+1}}.
\end{equation}
We thus have
\begin{equation}
\eta_{2n}\approx Re^{\frac{1}{\xi_{2n}-\xi_{2n+1}-1}},
\end{equation}
giving, for $n=1$, the well-known relation [19] for the dissipation scale $\eta_{2}\approx
Re^{\frac{1}{\xi_{2}-2}}$. Since by H\"older inequality, for all $q\geq p$, the exponents
$\xi_{p}\geq \xi_{q}$, it follows from (16) that $\eta_{2n}\geq \eta_{2n+1}$,
which justifies the derivation of expression (15).
By virtue of Eq.\ (6) we have
\begin{equation} \overline{(\frac{\partial u}{\partial
x})^{2n}}\propto Re^{\frac{\xi_{2n}-2n}{\xi_{2n}-\xi_{2n+1}-1}}.
\end{equation}
To compare Eq.\ (17) with the outcome of the multifractal formula (5),
we notice that both are the same in the limit $n\rightarrow
1$. In the limit $n \rightarrow \infty$, if the exponents
$\xi_{n}\rightarrow \xi_{\infty}=const$, or $\xi_{n}\rightarrow
\alpha n$, Eq.\ (17) and the outcome of mutifractal formula (5)
are identical.

The moments of velocity derivatives are given from Eq.\ (17) to be
\begin{equation}
S_{n}= \overline{(\frac{\partial u}{\partial
x})^{n}}/\overline{(\frac{\partial u}{\partial
x})^{2}}^{\frac{n}{2}} \propto Re^{s_{n}},
\end{equation}
with
\begin{equation} s_{2n}=
\frac{\xi_{2n}-2n}{\xi_{2n}-\xi_{2n+1}-1}-n.
\end{equation}
This expression can be evaluated readily if we know $\xi_{n}$.
Using the result obtained in [11,12] for $\xi_{n}$, we get

\begin{equation}
\xi_{n}\approx \frac{1.15n}{3(1+0.05n)}.
\end{equation}

The exponents $s_2$-$s_6$ from (19), after using (20), are listed
in Table I. Almost the same results are obtained if experimental
values [3] are chosen for $\xi_{n}$ instead of (20). These values
also agree very well with results from several phenomenological MF
models.

Several experimental measurements of $s_n$ are available in the
literature. For an incomplete list, see Ref.\ [18, 20, 21]. We
compare in Table I the theoretical numbers above with the data of
[21] in the atmospheric boundary layer at very high Reynolds
numbers and the latest wind tunnel measurements [18] in grid
turbulence. The differences between the two sets of experimental
numbers are a measure of uncertainty in the data. Keeping this in
mind, we may regard the agreement with the theoretical values to
be very good. The conclusions from other data sets [20] are quite
similar.

\begin{table}
\begin{center}
\begin{tabular}{|c|c|c|c|c|c|} \hline
exponent& $s_2$ & $s_3$ & $s_4$ & $s_5$ & $s_6$ \\ \hline
theory & 0 &
0.066 & 0.176 & 0.34 & 0.49 \\ \hline
experiment & 0 & 0.05 & 0.20 & 0.32 & 0.54 \\
& 0& (0.06) & (0.16) & (0.31) & (0.50) \\ \hline
\end{tabular}
\end{center}
\caption{The scaling exponents $s_n$ from theory and experiment.
The numbers in parentheses are from Ref.\ [18].}
\end{table}

As an aside, we note that the theory can also be used to show that
the results of Kolmogorov's Refined Similarity Hypotheses (RSH)
[15] are at least numerically close to the ones derived above. We
can evaluate the moments of velocity derivative
$\overline{(\partial_{x} u)^{2n}}$ by extrapolating RSH to $\eta$.
The dissipation averaged on this scale is of the same order as the
unaveraged dissipation. This assumption, commonly adopted in the
literature, appears reasonable if there is no structure for scales
smaller than $\eta$. From dissipation anomaly, we obtain
\begin{equation}
\epsilon_{\eta}\approx \epsilon\propto \frac{(\delta_{\eta}
u)^{3}}{\eta},
\end{equation}
where
\begin{equation}
\epsilon_{\eta}=\frac{1}{\eta^{3}}\int_{\eta}\epsilon(x)d^{3}x
\end{equation}
is the dissipation rate averaged over a ``ball'' of a radius $\eta$
with the center at ${\bf x}$. Equation (21) is an order of
magnitude estimate averaged over the ``universal'' Kolmogorov noise
$V=\eta\epsilon_{\eta}/(\delta_{\eta} u)^{3}$. Combining Eqs.\
(21) and (18) for the dissipation scale of the $3n$-th moment of
velocity difference we obtain
\begin{equation}
G_{2n}=\overline{(\frac{\partial u}{\partial x})^{2n}}\propto
Re^{g_{2n}},
\end{equation}
instead of Eq.\ (19), where
\begin{equation}
g_{2n}=n+\frac{\xi_{3n}-n}{\xi_{3n}-\xi_{3n+1}-1}.
\end{equation}
As we see, the two sets of formulae (18)-(19) and (23)-(24) are identical when
$n=1$ leading to the exact relation $\overline{(\partial_{x}
u)^{2}}\propto Re$. With the relation (20) for the scaling
exponents, both relations have the same asymptotics
$G_{2n}\rightarrow Re^{2n}$ in the limit $n\rightarrow \infty$. In
the interval $n\geq 1$, the formulae (18)-(19) and (23)-(24) differ by
no more than a few percent.

In summary, the theory developed here combines the exact equations
(7) and (8) with relations (9) and (10). Together, they lead to
Eqs.\ (18)-(19), which form the main result of the paper. While
this form is known from the MF theory, the present paper obtains
the exponents theoretically and the results agree well with
experiments.

It is useful to restate here the approximations involved in
derivation of (9) and (10). Expression (10) was obtained through
the model (9) for the acceleration terms in (7), and also as an
order-of-magnitude estimate from the equations for dissipation
anomaly. Since, at the dissipation scale $\eta$, the pressure
contribution simply renormalizes the coefficients in the left side
of equation (7), the expression (9) for the  viscous friction
force introduces $Re$-dependence into (7). This is the reason why,
for the fixed magnitude of the inertial range displacement $r$,
the structure functions $S_{n,m}(r)$ are $Re$-independent, while
the moments of derivatives are strongly $Re$-dependent. Though we
cannot prove that either scenario leading to (10) is rigorous, the
good agreement observed with experimental data gives us some
confidence that the theory is a step in the right direction.

We thank A. Bershadskii, T. Gotoh and A.M. Polyakov for helpful
discussions.


\begin{thebibliography}{99}
\bibitem{comment} Due to the spatial homogeneity of turbulence
$\overline{\delta a_{x}}=0$. This constraint is satisfied if, in
addition to (9), we account for the contribution from the
Bernoulli's equation-based model for the pressure gradient,
proposed in \cite{goto} and \cite{yakh2}, as
$$
\overline{\delta_{\eta} a_{x}|\delta_{\eta} u} \approx \frac{(\delta_{\eta}
u)^{3}}{\nu}+a\frac{(\delta_{\eta} u)^{2}}{\eta},
$$
where the constant $a$ is chosen from the condition that
$\frac{4}{5}Re=aA_{2}\eta_{2}^{-2+\xi_{2}}$. Here $A_{2}$ is the
amplitude in the relation $S_{2,0}(\eta)=A_{2}\eta_{2}^{\xi_{2}}$
(remembering that $\overline \epsilon = L = 1$). Since the
Kolmogorov constant $A_{2}\approx 2$ (see \cite{moni}), this gives
$a\approx 0.4$. The first term on the right side of the above
equation comes from viscous dissipation, while the second term
describes pressure effects on the dissipation scale $\eta$. This
term renormalizes the coefficients in front of the remaining
contributions to (7) and thus can be neglected in our analysis.
\end{thebibliography}
\end{document}